\documentclass{iau} 
\usepackage{graphicx}
\usepackage{natbib}
\usepackage{aas_macros}

\title[Nuclear Star Cluster Review] 
{The Properties of Nuclear Star Clusters and their Host Galaxies}

\author[Anil Seth, Nadine Neumayer, Torsten B{\"o}ker]   
{Anil Seth$^1$, Nadine Neumayer$^2$ 
 \and Torsten B{\"o}ker$^3$}

\affiliation{$^1$University of Utah, Salt Lake City \\ email: {\tt aseth@astro.utah.edu} \\[\affilskip]
  $^2$Max Planck Institute for Astronomy, Heidelberg \\[\affilskip]
$^3$European Space Agency, Space Telescope Science Center, Baltimore}

\pubyear{2019}
\volume{351}  
\jname{Star Clusters: From the Milky Way to the Early Universe}
\editors{A. Bragaglia, M.B. Davies, A. Sills \& E. Vesperini, eds.}
\begin{document}

\maketitle

\begin{abstract}
Nuclear star clusters are found at the centers of most galaxies.  They are the densest stellar systems in the Universe, and thus have unique and interesting stellar dynamics. We review how common nuclear star clusters are in galaxies of different masses and types, and then discuss the typical properties of NSCs.  We close by discussing the formation of NSCs, and how a picture is emerging of different formation mechanisms being dominant in lower and higher mass galaxies.

\end{abstract}

\firstsection 
\section{Introduction}

Nuclear star clusters (NSCs) are found at the centers of a majority of nearby spiral and elliptical galaxies.  NSCs are fascinating objects, and deserve further study for a wide range of reasons.  With millions of solar masses packed within the central few parsecs of a galaxy, NSCs are the densest stellar systems in the Universe and therfore host unique stellar dynamics.  They often co-exist with central black holes, and their stars can interact with the black hole to create spectacular tidal disruption events.  Unlike black holes, which provide no history of their formation, the stellar populations of an NSC provide a record of nuclear mass accretion.  Furthermore, when satellite galaxies are tidally stripped, their nuclei are generally unaffected, and thus stripped nuclei can provide information on the overall accretion history of galaxies as well.  These stripped nuclei appear to be hiding in plain view amongst the brightest globular clusters in our galaxy and others.  

The formation of NSCs is not yet fully understood, but two primary mechanisms are thought to be responsible: (1) the inspiral of globular star clusters to the center of the galaxy due to dynamical friction \citep[e.g.][]{tremaine75,lotz01,antonini13}, and/or (2) the {\em in situ} formation of stars from gas accreted into the galaxy center \citep[e.g.][]{mclaughlin06,hopkins10,brown18}.  

This contribution presents a brief review of NSCs drawn from an upcoming review article (Neumayer, Seth, B\"oker, {\em in prep for The Astronomy and Astrophysics Review}).  We urge readers to consult that article for a more complete review.  We start with discussing what galaxies host NSCs, and then discuss the basic properties of NSCs.  We conclude by highlighting an apparent transition in the dominant formation mechanism of NSCs at galaxy stellar masses of $\sim$10$^9$~M$_\odot$ from globular cluster accretion at lower masses to {\em in situ} star formation at higher masses.

\section{What Galaxies Have NSCs?}

\begin{figure}[h]
\begin{center}
  \includegraphics[height=2.4in]{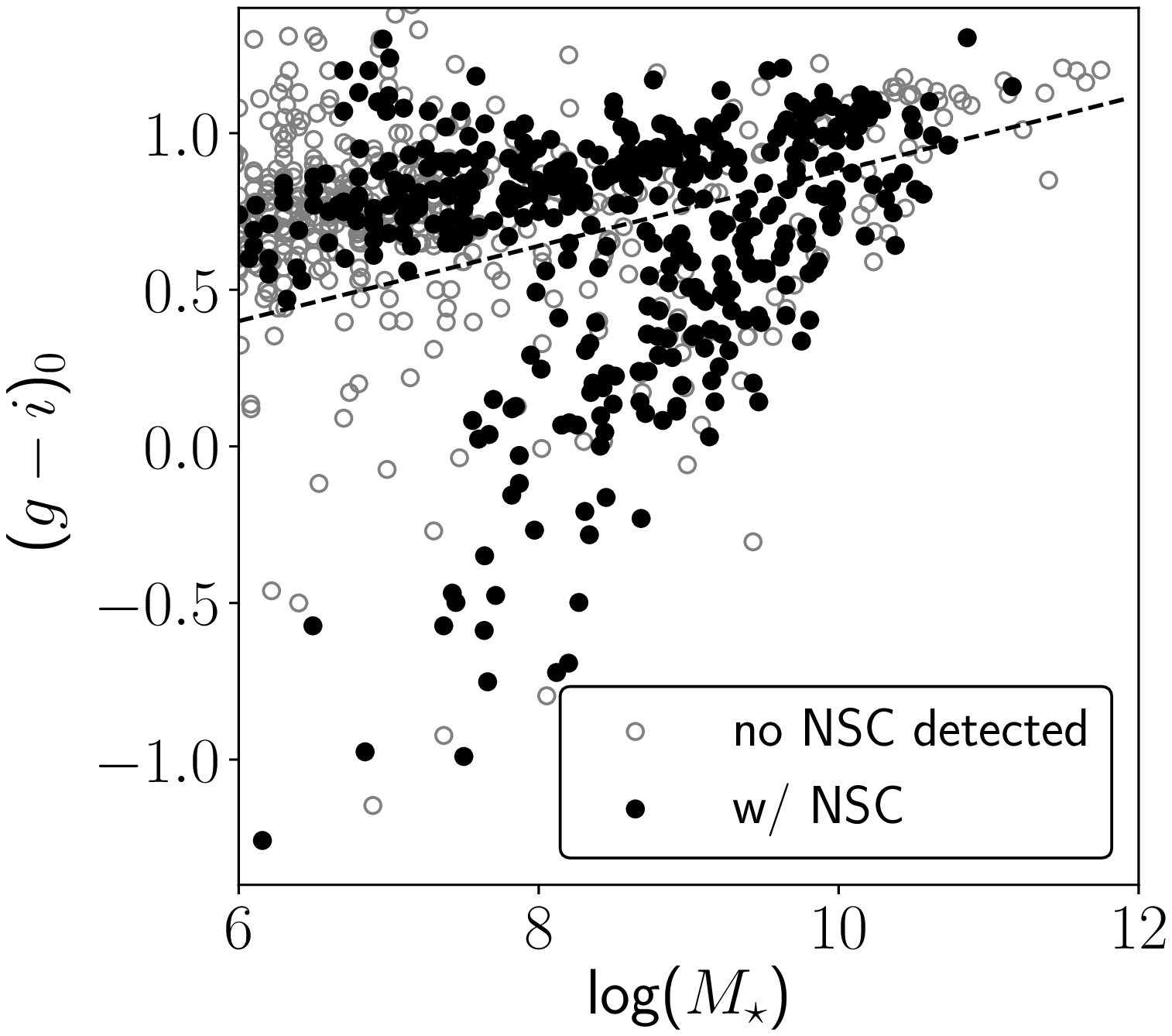}
  \includegraphics[height=2.4in]{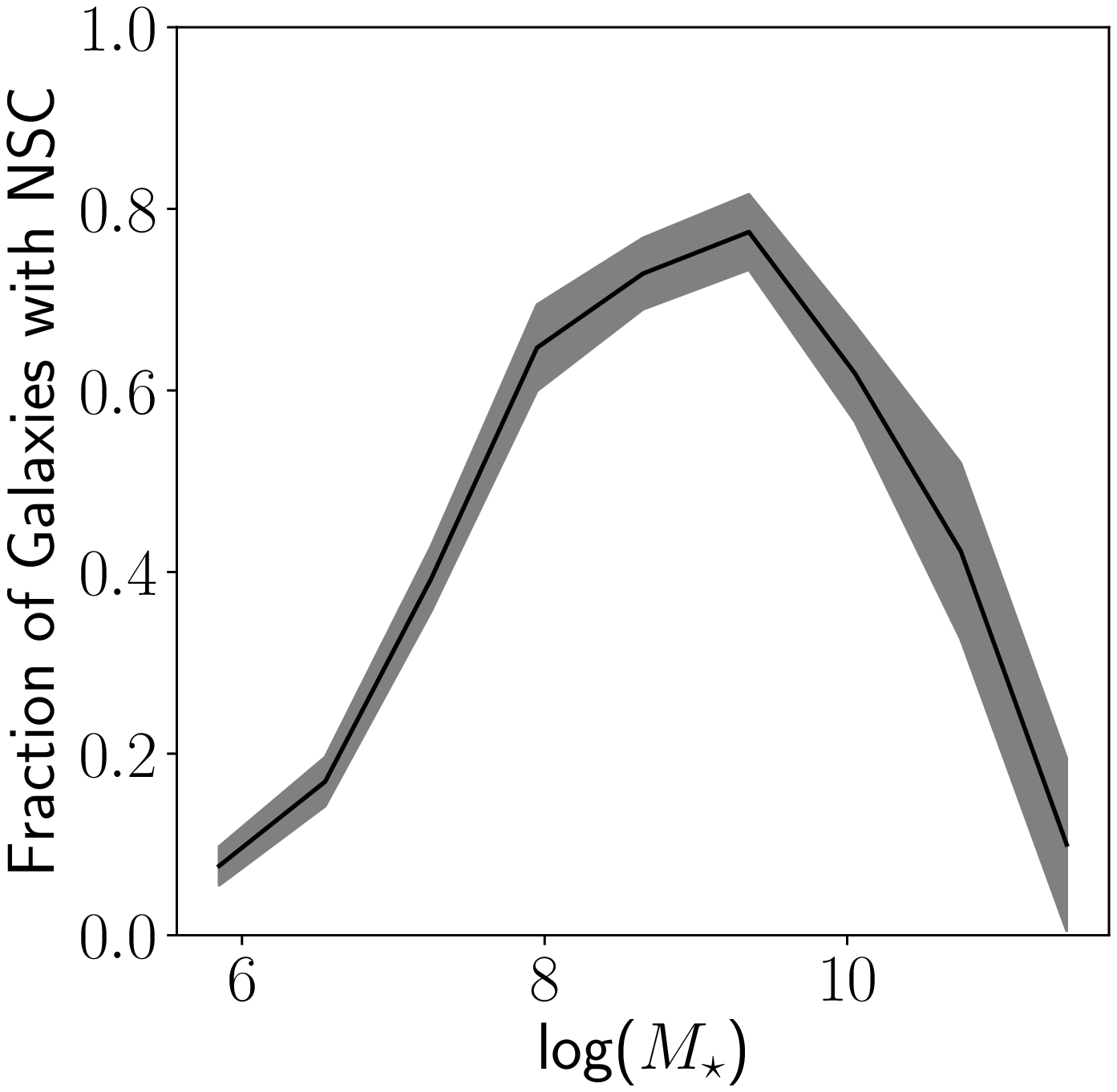} 
 \caption{The demographics of galaxies hosting NSCs. The plot combines data from studies of early-type galaxies in nearby clusters \citep{ferrarese06b,eigenthaler18,sanchez-janssen19}, and late-type field galaxies \citep{georgiev09,georgiev14}.  {\em Left --} Filled points indicate galaxies with clear NSCs, while open circles are galaxies where no NSC was identified.  For more details on the derivation of colors and stellar masses of the galaxies, see Neumayer, Seth \& B\"oker {\em in prep}. {\em Right --} The nucleation fraction as a function of galaxy mass shows a clear peak around $\sim$10$^9$~M$_\odot$.  This curve is similar to one presented in \citet{sanchez-janssen19}, but includes both early and late-type galaxies. We note that this represents a lower limit to the true nucleation fraction as in some galaxies nuclei were hard to identify due to dust extinction and confusion.  }
   \label{fig1}
\end{center}
\end{figure}

Early studies identified NSCs in nearby galaxies, including our own Milky Way and extending out to the Virgo cluster \citep{becklin68,light74,binggeli85}.  Thanks to large-scale nearby galaxy surveys with the Hubble Space Telescope \citep[e.g.][]{carollo97,boker02,cote06}, as well as deep ground-based surveys \citep[e.g.][]{ordenes-briceno18,sanchez-janssen19} we now have a much better census on how common NSCs are across all types of galaxies.  NSCs are typically easily identified as a clear overdensity in stars above the inwardly extrapolated light profile of a galaxy within the central 50~pc.  In most cases, NSCs are the brightest star clusters in their host galaxies.  However, we note that in some cases identifying or separating NSCs from the underlying galaxy is challenging due to the presence of dust, star formation, or multiple morphological structures.

In Fig.~1 we compile data from several surveys of nuclear star clusters in both early- and late-type galaxies.  In all cases, these surveys report on both the detection of NSCs and those galaxies in which they were not identified.  We caution that these non-detections can be due to dust obscuration or complexity and therefore do not necessarily imply no NSC is present.  The left panel shows that NSCs are ubiquitous in a wide range of galaxies from the red-sequence to the blue cloud, and especially in galaxies with stellar masses above $\sim$10$^8$~M$_\odot$.  The right panel shows the fraction of galaxies with NSCs making the trends with stellar mass easily visible.  Recent studies of NSCs in early-type galaxies in nearby galaxy clusters have clearly shown that the fraction of galaxies with NSCs declines at both the lowest and highest masses with a peak in the nucleation fraction at $\sim$10$^9$~M$_\odot$ \citep{cote06,denbrok14,sanchez-janssen19}.  A similar behavior is seen for late-type galaxies as well \citep{georgiev09,georgiev14}, although the turndown at the highest masses is less clear.

\begin{figure}[h]
\begin{center}
  \includegraphics[width=3.4in]{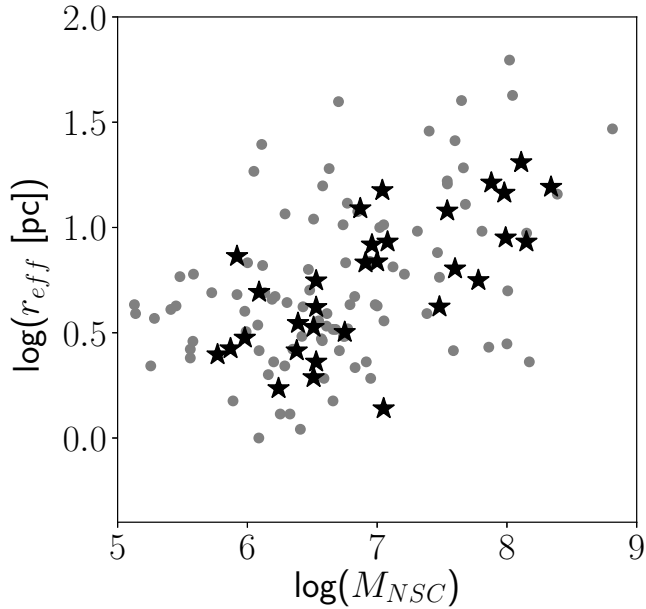}
 \caption{The mass and radii of NSCs in a wide range of galaxies.  Circles show photometric estimates of the NSC masses from \citet{georgiev16} and \citet{spengler17}, while the stars represent a compilation of dynamical and spectral synthesis mass estimates from \citet{erwin12}.}
   \label{fig1}
\end{center}
\end{figure}

\section{Properties of NSCs}

\noindent {\em Sizes, Masses, and Densities:} NSCs are compact and massive stellar systems with typical radii of $\sim$4~pc and masses ranging from $\sim$10$^{6-8}$~M$_\odot$ \citep[e.g.][]{georgiev16,spengler17}.  In Fig.~2 we plot the mass vs.~radius of NSCs in both early and late-type galaxies.  Below $\sim$10$^{7}$~M$_\odot$ there is almost no correlation of mass and radius, while above more massive NSCs are also typically larger in radius.  The densities of NSCs reach $\gtrsim$10$^6$~M$_\odot$/pc$^3$ in the central parsec \citep{lauer98}, and their densities correlates with galaxy stellar mass  (Pechetti et al., {\em in prep}).

\noindent {\em Stellar Populations:}  Unlike GCs, whose stars appear to have formed over a short period of time, NSCs clearly have multiple epochs of star formation.  This is most apparent in the two nearest NSCs at the center of our Milky Way and the Sgr dwarf galaxy, where stars can be resolved individually.  In both cases, their stars span a wide range of ages and metallicities \citep[e.g.][and Alfaro-Cuello in this proceedings]{siegel07,pfuhl11,feldmeier-krause15}.  In more distant galaxies, stellar population synthesis has also revealed that most NSCs seem to host stellar populations with a wide range of ages \citep[e.g.][]{walcher06,kacharov18}, with the youngest stars typically concentrated towards the center \citep{carson15}.  The metallicity of NSCs is also a key indicator of their formation, which we will discuss more in the next section.

\noindent {\em Morphology \& Kinematics:} Many NSCs are flattened, and in edge-on spiral galaxies, this flattening roughly aligns with the flattening of galaxies as a whole \citep{seth06}.  Rotation is also typically found in NSCs with $v/\sigma$ values of up to $\sim$1 being common \citep{seth08b,feldmeier14,nguyen18}.  

\noindent {\em Coincidence with Black Holes:}  There are now many examples of galaxies with both NSCs and central black holes, including our own Milky Way \citep[e.g.][]{seth08a,graham09,nguyen18}.  The lack of NSCs in the most massive early-type galaxies is likely due to binary BH merging \citep[e.g.][]{milosavljevic01}, although the presence of a BH can also hinder the formation NSCs through dynamical friction \citep{antonini13}.  There are also lower-mass galaxies where a NSC is present and a BH appears to be lacking \citep{gebhardt01,neumayer12}.  Thus far, no strong evidence for BHs has been found in galaxies below $\sim$10$^9$~M$_\odot$ \citep[e.g.][]{reines15}, a mass at which a vast majority of galaxies host NSCs (Figure~1).

\noindent {\em Correlation with Host Galaxy Properties:} The masses and luminosities of NSCs correlate with those of their host galaxies \citep[e.g.][]{balcells03,rossa06,ferrarese06}.  While initial studies suggested NSCs made up a constant fraction of a galaxy's stellar mass \citep{cote06}, it is now clear that this mass fraction varies with galaxy stellar mass.  Specifically, at the low mass end, the scaling of NSC mass with galaxy stellar mass is roughly $M_{NSC} \propto M_\star^{0.5}$ \citep[e.g.][]{scott13,sanchez-janssen19}, and thus an increasing fraction of a galaxies mass is present in an NSC at the lowest galaxy masses (reaching up to about 10\%); however note that this is accompanied by a decline in the fraction of galaxies that host NSCs at the lowest masses.  And while it was initially proposed that black holes and NSCs may share scaling relations with their hosts \citep{ferrarese06}, it is now clear that is not the case  \citep[e.g.][]{erwin12}.

\noindent {\em Stripped Nuclei and Ultracompact Dwarfs:} During galaxy interactions, satellite galaxies can be fully stripped of their envelopes leaving behind only their NSCs \citep{bekki01a,pfeffer13}; these remnants may be hiding amongs the most massive globular clusters and ultracompact dwarf galaxies \citep{drinkwater03}.  The Sgr dwarf galaxy is an example of a snapshot in this process \citep{ibata94}, with its NSC (the globular cluster M54) being the first part of the galaxy to be detected.  The detection of supermassive black holes in the most massive ultracompact dwarf galaxies has provided strong evidence that they are in fact stripped NSCs \citep[e.g.][]{mieske13,seth14,ahn17}.  These stripped nuclei are as common in clusters as present day nuclei, and may contain up to a third of the massive black holes in the local Universe \citep{voggel19}.  We note that an ongoing challenge is distinguishing stripped nuclei from ordinary globular clusters in cases where significant age spreads or a massive black hole are not present.  Recent theoretical work has predicted a range of stripped nuclei in the Milky Way from $\sim$1-6 \citep{pfeffer14,kruijssen19}.  The number of Milky Way globular clusters with metallicity spreads is now quite a bit larger than this expected number of stripped nuclei, suggesting perhaps that metallicity spreads alone are not sufficient to identify a stripped nucleus \citep[e.g.][]{dacosta16,marino18}.

\begin{figure}[h]
\begin{center}
  \includegraphics[width=3.4in]{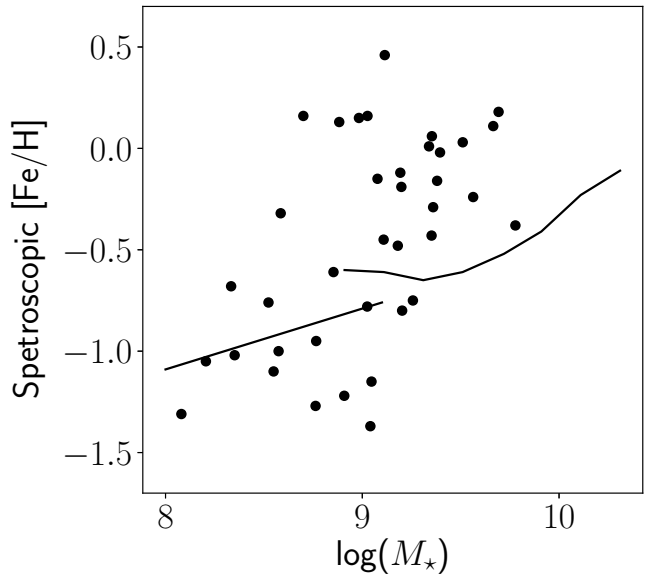}
 \caption{Spectroscopic metallicity estimates of NSCs in early-type galaxies show a change with galaxy stellar mass \citep{koleva09,paudel11}.  At low galaxy masses, the NSC metallicities often fall below the galaxy mass-metallicity relation \citep[solid lines;][]{gallazzi05,kirby13}, while at higher galaxy masses almost all are above. }
   \label{fig1}
\end{center}
\end{figure}

\section{An Emerging Picture of the Mass-Dependence of NSC Formation}

As noted in the introduction, two main scenarios are thought to be responsible for NSC formation: (1) globular cluster inspiral due to dynamical friction, and (2) {\em in situ} star formation from gas that reaches the nucleus.  From recent literature, it is becoming increasingly clear that the dominant formation mechanism changes with galaxy mass, with a transition happening around stellar masses of 10$^9$~M$_\odot$.  This transition mass represents an upturn in the scaling relations between the luminosity/mass of NSCs vs. those of their galaxy \citep{denbrok14,sanchez-janssen19}.  At low-masses the $M_{NSC} \propto M_\star^{0.5}$ scaling meets the expectations of globular cluster inspiral \citep{gnedin14}.  The densities of NSCs are also found to be significantly higher than typical globular clusters in galaxies with stellar masses above 10$^9$~M$_\odot$, suggesting {\em in situ} star formation may be playing an important role (Pechetti et al., {\em in prep}).  The important role of ongoing star formation in galaxies above 10$^9$~M$_\odot$ is also supported by the observations of young stars in many of these galaxies \citep[e.g.][]{walcher06}.

The metallicity of NSCs is a particularly revealing quantity for determining their formation, as the inspiral model would create the possibility of having more metal-poor stars at the center than in the surrounding galaxy.  In contrast, the accretion of polluted gas into the center would suggest that stars will equal or higher metallicity as their surrounding stars \citep[e.g.][]{brown18}.  Figure~3 shows spectroscopic metallicities in early-type NSCs from two papers \citep{koleva09,paudel11}, compared to the mass-metallicty relationship of galaxies \citep{gallazzi05,kirby13}.  A clear transition is visible $\sim$10$^9$~M$_\odot$ where above this mass all NSCs are more metal-rich than the typical galaxy, while below this mass a wider range of metallicities is seen, with some NSCs significantly more metal-poor than the typical galaxy.  A low metallicity ([Fe/H]$\sim -1.6$) is found for the dominant population in M54 \citep{carretta10}, whose stellar mass is estimated to be a few time 10$^8$~M$_\odot$.  Several NSCs around Cen~A are also found to be more metal-poor than their surrounding stars (Seth et al. {\em in prep}).  Overall, the existing metallicity measurements clearly support a scenario where low mass galaxies form their NSCs primarily through GC accretion, while at high masses {\em in situ} star formation becomes more important.  These observations raise the unsolved question of why there is a transition at 10$^9$~M$_\odot$, and whether this might be related to the increasing importance of black holes at these higher masses.

{\em Acknowledgments:} AS acknowledges funding for this work from NSF grants AST-1350389 \& AST-1813609.


\begin{thebibliography}{61}
\expandafter\ifx\csname natexlab\endcsname\relax\def\natexlab#1{#1}\fi

\bibitem[{{Ahn} {et~al.}(2017){Ahn}, {Seth}, {den Brok}, {Strader},
  {Baumgardt}, {van den Bosch}, {Chilingarian}, {Frank}, {Hilker}, {McDermid},
  {Mieske}, {Romanowsky}, {Spitler}, {Brodie}, {Neumayer}, \& {Walsh}}]{ahn17}
{Ahn}, C.~P. {et~al.} 2017, \apj, 839, 72

\bibitem[{{Antonini}(2013)}]{antonini13}
{Antonini}, F. 2013, \apj, 763, 62

\bibitem[{{Balcells} {et~al.}(2003){Balcells}, {Graham},
  {Dom{\'\i}nguez-Palmero}, \& {Peletier}}]{balcells03}
{Balcells}, M., {Graham}, A.~W., {Dom{\'\i}nguez-Palmero}, L., \& {Peletier},
  R.~F. 2003, \apj, 582, L79

\bibitem[{{Becklin} \& {Neugebauer}(1968)}]{becklin68}
{Becklin}, E.~E., \& {Neugebauer}, G. 1968, \apj, 151, 145

\bibitem[{{Bekki} {et~al.}(2001){Bekki}, {Couch}, \& {Drinkwater}}]{bekki01a}
{Bekki}, K., {Couch}, W.~J., \& {Drinkwater}, M.~J. 2001, \apjl, 552, L105

\bibitem[{{Binggeli} {et~al.}(1985){Binggeli}, {Sandage}, \&
  {Tammann}}]{binggeli85}
{Binggeli}, B., {Sandage}, A., \& {Tammann}, G.~A. 1985, \aj, 90, 1681

\bibitem[{{B{\"o}ker} {et~al.}(2002){B{\"o}ker}, {Laine}, {van der Marel},
  {Sarzi}, {Rix}, {Ho}, \& {Shields}}]{boker02}
{B{\"o}ker}, T. {et~al.} 2002, \aj, 123, 1389

\bibitem[{{Brown} {et~al.}(2018){Brown}, {Gnedin}, \& {Li}}]{brown18}
{Brown}, G., {Gnedin}, O.~Y., \& {Li}, H. 2018, \apj, 864, 94

\bibitem[{{Carollo} {et~al.}(1997){Carollo}, {Stiavelli}, {de Zeeuw}, \&
  {Mack}}]{carollo97}
{Carollo}, C.~M., {Stiavelli}, M., {de Zeeuw}, P.~T., \& {Mack}, J. 1997, \aj,
  114, 2366

\bibitem[{{Carretta} {et~al.}(2010){Carretta}, {Bragaglia}, {Gratton},
  {Lucatello}, {Bellazzini}, {Catanzaro}, {Leone}, {Momany}, {Piotto}, \&
  {D'Orazi}}]{carretta10}
{Carretta}, E. {et~al.} 2010, \aap, 520, A95

\bibitem[{{Carson} {et~al.}(2015){Carson}, {Barth}, {Seth}, {den Brok},
  {Cappellari}, {Greene}, {Ho}, \& {Neumayer}}]{carson15}
{Carson}, D.~J. {et~al.} 2015, \aj, 149, 170

\bibitem[{{C{\^o}t{\'e}} {et~al.}(2006){C{\^o}t{\'e}}, {Piatek}, {Ferrarese},
  {Jord{\'a}n}, {Merritt}, {Peng}, {Ha{\c s}egan}, {Blakeslee}, {Mei}, {West},
  {Milosavljevi{\'c}}, \& {Tonry}}]{cote06}
{C{\^o}t{\'e}}, P. {et~al.} 2006, \apjs, 165, 57

\bibitem[{{Da Costa}(2016)}]{dacosta16}
{Da Costa}, G.~S. 2016, in IAU Symposium, Vol. 317, The General Assembly of
  Galaxy Halos: Structure, Origin and Evolution, ed. A.~{Bragaglia},
  M.~{Arnaboldi}, M.~{Rejkuba}, \& D.~{Romano}, 110--115

\bibitem[{{den Brok} {et~al.}(2014){den Brok}, {Peletier}, {Seth}, {Balcells},
  {Dominguez}, {Graham}, {Carter}, {Erwin}, {Ferguson}, {Goudfrooij},
  {Guzm{\'a}n}, {Hoyos}, {Jogee}, {Lucey}, {Phillipps}, {Puzia}, {Valentijn},
  {Kleijn}, \& {Weinzirl}}]{denbrok14}
{den Brok}, M. {et~al.} 2014, \mnras, 445, 2385

\bibitem[{{Drinkwater} {et~al.}(2003){Drinkwater}, {Gregg}, {Hilker}, {Bekki},
  {Couch}, {Ferguson}, {Jones}, \& {Phillipps}}]{drinkwater03}
{Drinkwater}, M.~J. {et~al.} 2003, \nat, 423, 519

\bibitem[{{Eigenthaler} {et~al.}(2018){Eigenthaler}, {Puzia}, {Taylor},
  {Ordenes-Brice{\~n}o}, {Mu{\~n}oz}, {Ribbeck}, {Alamo-Mart{\'\i}nez},
  {Zhang}, {{\'A}ngel}, \& {Capaccioli}}]{eigenthaler18}
{Eigenthaler}, P. {et~al.} 2018, \apj, 855, 142

\bibitem[{{Erwin} \& {Gadotti}(2012)}]{erwin12}
{Erwin}, P., \& {Gadotti}, D.~A. 2012, Advances in Astronomy, 2012, 946368

\bibitem[{{Feldmeier} {et~al.}(2014){Feldmeier}, {Neumayer}, {Seth},
  {Sch{\"o}del}, {L{\"u}tzgendorf}, {de Zeeuw}, {Kissler-Patig}, {Nishiyama},
  \& {Walcher}}]{feldmeier14}
{Feldmeier}, A. {et~al.} 2014, \aap, 570, A2

\bibitem[{{Feldmeier-Krause} {et~al.}(2015){Feldmeier-Krause}, {Neumayer},
  {Sch{\"o}del}, {Seth}, {Hilker}, {de Zeeuw}, {Kuntschner}, {Walcher},
  {L{\"u}tzgendorf}, \& {Kissler-Patig}}]{feldmeier-krause15}
{Feldmeier-Krause}, A. {et~al.} 2015, \aap, 584, A2

\bibitem[{{Ferrarese} {et~al.}(2006{\natexlab{a}}){Ferrarese}, {C{\^o}t{\'e}},
  {Dalla Bont{\`a}}, {Peng}, {Merritt}, {Jord{\'a}n}, {Blakeslee}, {Ha{\c
  s}egan}, {Mei}, {Piatek}, {Tonry}, \& {West}}]{ferrarese06}
{Ferrarese}, L. {et~al.} 2006{\natexlab{a}}, \apjl, 644, L21

\bibitem[{{Ferrarese} {et~al.}(2006{\natexlab{b}}){Ferrarese}, {C{\^o}t{\'e}},
  {Jord{\'a}n}, {Peng}, {Blakeslee}, {Piatek}, {Mei}, {Merritt},
  {Milosavljevi{\'c}}, \& {Tonry}}]{ferrarese06b}
---. 2006{\natexlab{b}}, \apjs, 164, 334

\bibitem[{{Gallazzi} {et~al.}(2005){Gallazzi}, {Charlot}, {Brinchmann},
  {White}, \& {Tremonti}}]{gallazzi05}
{Gallazzi}, A. {et~al.} 2005, \mnras, 362, 41

\bibitem[{{Gebhardt} {et~al.}(2001){Gebhardt}, {Lauer}, {Kormendy}, {Pinkney},
  {Bower}, {Green}, {Gull}, {Hutchings}, {Kaiser}, \& {Nelson}}]{gebhardt01}
{Gebhardt}, K. {et~al.} 2001, \aj, 122, 2469

\bibitem[{{Georgiev} \& {B{\"o}ker}(2014)}]{georgiev14}
{Georgiev}, I.~Y., \& {B{\"o}ker}, T. 2014, \mnras, 441, 3570

\bibitem[{{Georgiev} {et~al.}(2016){Georgiev}, {B{\"o}ker}, {Leigh},
  {L{\"u}tzgendorf}, \& {Neumayer}}]{georgiev16}
{Georgiev}, I.~Y. {et~al.} 2016, \mnras, 457, 2122

\bibitem[{{Georgiev} {et~al.}(2009){Georgiev}, {Hilker}, {Puzia}, {Goudfrooij},
  \& {Baumgardt}}]{georgiev09}
---. 2009, \mnras, 396, 1075

\bibitem[{{Gnedin} {et~al.}(2014){Gnedin}, {Ostriker}, \&
  {Tremaine}}]{gnedin14}
{Gnedin}, O.~Y., {Ostriker}, J.~P., \& {Tremaine}, S. 2014, \apj, 785, 71

\bibitem[{{Graham} \& {Spitler}(2009)}]{graham09}
{Graham}, A.~W., \& {Spitler}, L.~R. 2009, \mnras, 397, 2148

\bibitem[{{Hopkins} \& {Quataert}(2010)}]{hopkins10}
{Hopkins}, P.~F., \& {Quataert}, E. 2010, \mnras, 405, L41

\bibitem[{{Ibata} {et~al.}(1994){Ibata}, {Gilmore}, \& {Irwin}}]{ibata94}
{Ibata}, R.~A., {Gilmore}, G., \& {Irwin}, M.~J. 1994, \nat, 370, 194

\bibitem[{{Kacharov} {et~al.}(2018){Kacharov}, {Neumayer}, {Seth},
  {Cappellari}, {McDermid}, {Walcher}, \& {B{\"o}ker}}]{kacharov18}
{Kacharov}, N. {et~al.} 2018, \mnras, 480, 1973

\bibitem[{{Kirby} {et~al.}(2013){Kirby}, {Cohen}, {Guhathakurta}, {Cheng},
  {Bullock}, \& {Gallazzi}}]{kirby13}
{Kirby}, E.~N. {et~al.} 2013, \apj, 779, 102

\bibitem[{{Koleva} {et~al.}(2009){Koleva}, {de Rijcke}, {Prugniel},
  {Zeilinger}, \& {Michielsen}}]{koleva09}
{Koleva}, M. {et~al.} 2009, \mnras, 396, 2133

\bibitem[{{Kruijssen} {et~al.}(2019){Kruijssen}, {Pfeffer}, {Reina-Campos},
  {Crain}, \& {Bastian}}]{kruijssen19}
{Kruijssen}, J.~M.~D. {et~al.} 2019, \mnras, 486, 3180

\bibitem[{{Lauer} {et~al.}(1998){Lauer}, {Faber}, {Ajhar}, {Grillmair}, \&
  {Scowen}}]{lauer98}
{Lauer}, T.~R. {et~al.} 1998, \aj, 116, 2263

\bibitem[{{Light} {et~al.}(1974){Light}, {Danielson}, \&
  {Schwarzschild}}]{light74}
{Light}, E.~S., {Danielson}, R.~E., \& {Schwarzschild}, M. 1974, \apj, 194, 257

\bibitem[{{Lotz} {et~al.}(2001){Lotz}, {Telford}, {Ferguson}, {Miller},
  {Stiavelli}, \& {Mack}}]{lotz01}
{Lotz}, J.~M. {et~al.} 2001, \apj, 552, 572

\bibitem[{{Marino} {et~al.}(2018){Marino}, {Yong}, {Milone}, {Piotto},
  {Lundquist}, {Bedin}, {Chen{\'e}}, {Da Costa}, {Asplund}, \&
  {Jerjen}}]{marino18}
{Marino}, A.~F. {et~al.} 2018, \apj, 859, 81

\bibitem[{{McLaughlin} {et~al.}(2006){McLaughlin}, {King}, \&
  {Nayakshin}}]{mclaughlin06}
{McLaughlin}, D.~E., {King}, A.~R., \& {Nayakshin}, S. 2006, \apjl, 650, L37

\bibitem[{{Mieske} {et~al.}(2013){Mieske}, {Frank}, {Baumgardt},
  {L{\"u}tzgendorf}, {Neumayer}, \& {Hilker}}]{mieske13}
{Mieske}, S. {et~al.} 2013, \aap, 558, A14

\bibitem[{{Milosavljevi{\'c}} \& {Merritt}(2001)}]{milosavljevic01}
{Milosavljevi{\'c}}, M., \& {Merritt}, D. 2001, \apj, 563, 34

\bibitem[{{Neumayer} \& {Walcher}(2012)}]{neumayer12}
{Neumayer}, N., \& {Walcher}, C.~J. 2012, Advances in Astronomy, 2012, 709038

\bibitem[{{Nguyen} {et~al.}(2018){Nguyen}, {Seth}, {Neumayer}, {Kamann},
  {Voggel}, {Cappellari}, {Picotti}, {Nguyen}, {B{\"o}ker}, {Debattista},
  {Caldwell}, {McDermid}, {Bastian}, {Ahn}, \& {Pechetti}}]{nguyen18}
{Nguyen}, D.~D. {et~al.} 2018, \apj, 858, 118

\bibitem[{{Ordenes-Brice{\~n}o} {et~al.}(2018){Ordenes-Brice{\~n}o}, {Puzia},
  {Eigenthaler}, {Taylor}, {Mu{\~n}oz}, {Zhang}, {Alamo-Mart{\'{\i}}nez},
  {Ribbeck}, {Grebel}, {{\'A}ngel}, {C{\^o}t{\'e}}, {Ferrarese}, {Hilker},
  {Lan{\c c}on}, {Mieske}, {Miller}, {Rong}, \&
  {S{\'a}nchez-Janssen}}]{ordenes-briceno18}
{Ordenes-Brice{\~n}o}, Y. {et~al.} 2018, \apj, 860, 4

\bibitem[{{Paudel} {et~al.}(2011){Paudel}, {Lisker}, \&
  {Kuntschner}}]{paudel11}
{Paudel}, S., {Lisker}, T., \& {Kuntschner}, H. 2011, \mnras, 413, 1764

\bibitem[{{Pfeffer} \& {Baumgardt}(2013)}]{pfeffer13}
{Pfeffer}, J., \& {Baumgardt}, H. 2013, \mnras, 433, 1997

\bibitem[{{Pfeffer} {et~al.}(2014){Pfeffer}, {Griffen}, {Baumgardt}, \&
  {Hilker}}]{pfeffer14}
{Pfeffer}, J., {Griffen}, B.~F., {Baumgardt}, H., \& {Hilker}, M. 2014, \mnras,
  444, 3670

\bibitem[{{Pfuhl} {et~al.}(2011){Pfuhl}, {Fritz}, {Zilka}, {Maness},
  {Eisenhauer}, {Genzel}, {Gillessen}, {Ott}, {Dodds-Eden}, \&
  {Sternberg}}]{pfuhl11}
{Pfuhl}, O. {et~al.} 2011, \apj, 741, 108

\bibitem[{{Reines} \& {Volonteri}(2015)}]{reines15}
{Reines}, A.~E., \& {Volonteri}, M. 2015, \apj, 813, 82

\bibitem[{{Rossa} {et~al.}(2006){Rossa}, {van der Marel}, {B{\"o}ker},
  {Gerssen}, {Ho}, {Rix}, {Shields}, \& {Walcher}}]{rossa06}
{Rossa}, J. {et~al.} 2006, \aj, 132, 1074

\bibitem[{{S{\'a}nchez-Janssen} {et~al.}(2019){S{\'a}nchez-Janssen},
  {C{\^o}t{\'e}}, {Ferrarese}, {Peng}, {Roediger}, {Blakeslee}, {Emsellem},
  {Puzia}, {Spengler}, \& {Taylor}}]{sanchez-janssen19}
{S{\'a}nchez-Janssen}, R. {et~al.} 2019, \apj, 878, 18

\bibitem[{{Scott} \& {Graham}(2013)}]{scott13}
{Scott}, N., \& {Graham}, A.~W. 2013, \apj, 763, 76

\bibitem[{{Seth} {et~al.}(2008{\natexlab{a}}){Seth}, {Ag{\"u}eros}, {Lee}, \&
  {Basu-Zych}}]{seth08a}
{Seth}, A., {Ag{\"u}eros}, M., {Lee}, D., \& {Basu-Zych}, A.
  2008{\natexlab{a}}, \apj, 678, 116

\bibitem[{{Seth} {et~al.}(2008{\natexlab{b}}){Seth}, {Blum}, {Bastian},
  {Caldwell}, \& {Debattista}}]{seth08b}
{Seth}, A.~C. {et~al.} 2008{\natexlab{b}}, \apj, 687, 997

\bibitem[{{Seth} {et~al.}(2006){Seth}, {Dalcanton}, {Hodge}, \&
  {Debattista}}]{seth06}
{Seth}, A.~C., {Dalcanton}, J.~J., {Hodge}, P.~W., \& {Debattista}, V.~P. 2006,
  \aj, 132, 2539

\bibitem[{{Seth} {et~al.}(2014){Seth}, {van den Bosch}, {Mieske}, {Baumgardt},
  {Brok}, {Strader}, {Neumayer}, {Chilingarian}, {Hilker}, {McDermid},
  {Spitler}, {Brodie}, {Frank}, \& {Walsh}}]{seth14}
{Seth}, A.~C. {et~al.} 2014, \nat, 513, 398

\bibitem[{{Siegel} {et~al.}(2007){Siegel}, {Dotter}, {Majewski}, {Sarajedini},
  {Chaboyer}, {Nidever}, {Anderson}, {Mar{\'{\i}}n-Franch}, {Rosenberg},
  {Bedin}, {Aparicio}, {King}, {Piotto}, \& {Reid}}]{siegel07}
{Siegel}, M.~H. {et~al.} 2007, \apjl, 667, L57

\bibitem[{{Spengler} {et~al.}(2017){Spengler}, {C{\^o}t{\'e}}, {Roediger},
  {Ferrarese}, {S{\'a}nchez-Janssen}, {Toloba}, {Liu}, {Guhathakurta},
  {Cuillandre}, {Gwyn}, {Zirm}, {Mu{\~n}oz}, {Puzia}, {Lan{\c{c}}on}, {Peng},
  {Mei}, \& {Powalka}}]{spengler17}
{Spengler}, C. {et~al.} 2017, \apj, 849, 55

\bibitem[{{Tremaine} {et~al.}(1975){Tremaine}, {Ostriker}, \&
  {Spitzer}}]{tremaine75}
{Tremaine}, S.~D., {Ostriker}, J.~P., \& {Spitzer}, L. 1975, \apj, 196, 407

\bibitem[{{Voggel} {et~al.}(2019){Voggel}, {Seth}, {Baumgardt}, {Mieske},
  {Pfeffer}, \& {Rasskazov}}]{voggel19}
{Voggel}, K.~T. {et~al.} 2019, \apj, 871, 159

\bibitem[{{Walcher} {et~al.}(2006){Walcher}, {B{\"o}ker}, {Charlot}, {Ho},
  {Rix}, {Rossa}, {Shields}, \& {van der Marel}}]{walcher06}
{Walcher}, C.~J. {et~al.} 2006, \apj, 649, 692

\end{thebibliography}

\end{document}